\documentclass[12pt]{article}

\renewcommand{\thefootnote}{\fnsymbol{footnote}}
\begin{document}

\title{\ \\*[10pt]
{\bf \Large 
Sequestering in models of F-term uplifting\footnote{
Talks given at PASCOS08 
(the 14th International Symposium on Particles, 
Strings and Cosmology, 
Perimeter Institute, Waterloo, Canada, June 2-6, 2008) 
and SUSY08 
(the 16th International Conference on Supersymmetry
and the Unification of Fundamental Interactions, 
the COEX Center, Seoul, Korea, June 16-21, 2008). 
}}\\*[20pt]
}

\author{Hiroyuki~Abe\footnote{
E-mail address: abe@yukawa.kyoto-u.ac.jp} \\*[20pt]
{\it \normalsize 
Yukawa Institute for Theoretical Physics, 
Kyoto University, Kyoto 606-8502, Japan} \\*[50pt]}

\date{
\centerline{\small \bf Abstract}
\begin{minipage}{0.9\linewidth}
\medskip 
\medskip 
\small
We discuss the nature of sequestering supersymmetry breaking 
sectors in a certain class of moduli stabilization in 
supergravity/string models, where a negative vacuum energy 
of the nonperturbative moduli potential is canceled by 
dynamically generated F-terms. 
Two illustrating examples are shown to sketch the issues 
around the supersymmetry breaking, flavors and sequestering 
within such a framework. 
\end{minipage}
}

\begin{titlepage}
\maketitle
\thispagestyle{empty}
\end{titlepage}

\renewcommand{\thefootnote}{\arabic{footnote}}
\setcounter{footnote}{0}

\section{Introduction}
The low-scale supersymmetry (SUSY) is one of the best solutions 
to the weak-Planck hierarchy problem, and then a good candidate 
for the physics at the TeV scale beyond the standard model (SM). 
The minimal SUSY SM (MSSM) unifies three gauge coupling 
constants at a certain high energy scale, and provides a dark matter 
candidate as the lightest SUSY particle. The existence of SUSY is also 
predicted by the only known consistent framework of quantum gravity, 
i.e., superstring theory. 

In constructing realistic SUSY models, there are several conditions 
to be satisfied. First of all, in order to avoid the so-called 
supertrace theorem, the SUSY would be broken in the hidden sector 
and communicated to the visible (MSSM) sector by some messenger fields. 
The low-energy SUSY particle spectrum is determined by the mediation 
mechanism. One of the severest constraints on the mediation structure 
comes from the observations of flavor changing processes, 
i.e., SUSY flavor violations. 

The hidden sector should not couple to the visible sector, by definition, 
through tree-level renormalizable interactions. Such a structure would 
be naturally realized in a higher-dimensional spacetime, where both sectors 
are somehow separated from each other geometrically in extra dimensions~\cite{
Randall:1998uk}, which is sometimes referred to as {\it sequestering}. 
The higher-dimensional spacetime is also predicted by string theory. 
Aside from the anomaly mediation~\cite{Randall:1998uk,Giudice:1998xp}, 
the most natural candidate for the messenger in such a situation is the 
moduli which govern the size and the shape of the extra dimensions. 
A more model-dependent candidate is SM-charged massive fields if they exist, 
i.e., the gauge mediation. 

Because moduli originate from the higher-dimensional gravity multiplet, the 
dominant interactions with the visible sector as well as with the hidden sector 
appear as tree-level nonrenormalizable terms, which are determined by the 
geometry of the extra dimensions. The moduli, as it stands, are flat-directions 
of the tree-level or the perturbative potential, which should be stabilized by 
some mechanisms such as nonperturbative dynamics. The magnitude of the SUSY 
breaking mediated by moduli is also determined by the stabilization mechanism. 
One of the most challenging issues around the moduli stabilization is a 
realization of an almost vanishing vacuum energy at a minimum of the moduli 
potential, required by the observation of our universe.

\section{Uplifting}
Most of $N=1$ compactifications of supergravity/string have moduli 
in the low-energy effective theory. 
Many of them can be stabilized by turning on some fluxes/torsions 
in extra-dimensional space, while some of them can not. 
The remaining moduli would be stabilized by field theoretical 
nonperturbative effects such as gaugino condensations. 
However, it is known that such effects generically yield 
negative vacuum energies at the minima of moduli potential, 
none of which corresponds to our universe. 

There has been proposed a systematic way for fixing the moduli at a 
supersymmetry breaking minimum with a vanishing vacuum energy~\cite{
Kachru:2003aw}, decoupling the SUSY breaking sector from the moduli 
as well as from the MSSM sector. In this case, the SUSY breaking (hidden) 
sector is also called an {\it uplifting} sector, because the SUSY 
breaking energy lifts the negative energy of the nonperturbative 
moduli potential to be vanishing. A simple and natural candidate for 
the source of the uplifting energy is the F-terms of dynamical SUSY 
breaking sectors. We call this scenario the F-term uplifting. 
It is notable that there would exist a large number of (metastable) 
SUSY breaking states in supergravity/string models caused by, e.g., 
SUSY gauge dynamics~\cite{Intriligator:2006dd}, that can provide 
the uplifting F-terms in a controllable manner (
see, e.g., Ref~\cite{Dudas:2006gr,Abe:2006xp}).

\section{Low energy effective theory}
The relevant quantities to our discussion in a generic four-dimensional 
(4D) $N=1$ effective supergravity are 
\begin{eqnarray}
\Omega \!\!\!&=&\!\!\! 
\Omega_0(T,\bar{T})
+Y_{X \bar{X}}(T,\bar{T})|X|^2
+Y_{I\bar{J}}(T,\bar{T},|X|^2)Q^IQ^{\bar{J}}
\nonumber \\ && \qquad 
+L_{IJ}(T,\bar{T},|X|^2)Q^IQ^J+{\rm h.c.}
+\cdots, 
\nonumber \\
W \!\!\!&=&\!\!\! 
W_0(T)
+f(T)X
+\mu_{IJ}(T) Q^I Q^J/2!
+\lambda_{IJK}(T)Q^IQ^JQ^K/3!
+\cdots, 
\nonumber \\
f_a \!\!\!&=&\!\!\! l_a+k_aT+\cdots, 
\nonumber
\end{eqnarray}
where $K=-3 \ln (-\Omega/3)$ is the K\"ahler potential, 
$W$ is the superpotential and 
$f_a$ are the gauge kinetic functions 
determining the gauge couplings associated 
with vector multiplets labelled by the index $a$ 
as $\langle {\rm Re}f_a \rangle=1/g_a^2$. 
The chiral multiplets $T$, $X$ and $Q^I$ represent 
the light moduli, 
the uplifting (SUSY breaking) 
and the visible (MSSM) sectors, respectively. 
The ellipses denote irrelevant or higher-order terms in powers 
of $X$ and $Q^I$, some of which might be suppressed due to some 
(approximate) symmetries, such as an $U(1)_R$-symmetry which is 
responsible for the dynamical SUSY breaking and assumed to be 
broken by $W_0$~\cite{Intriligator:2007py}. 

From these functions, we can evaluate soft SUSY breaking terms like 
the gaugino masses $M_a$, 
the scalar masses $m^2_I$, 
and the scalar trilinear couplings $A_{IJK}$ (A-terms) 
in the MSSM sector as 
\begin{eqnarray}
M_a \!\!\!&=&\!\!\! F^i \partial_i \ln ({\rm Re } f_a), 
\nonumber \\
m^2_I \!\!\!&=&\!\!\! 
-F^i \bar{F}^{\bar{j}} 
\partial_i \partial_{\bar{j}} \ln Y_{I\bar{I}}, 
\nonumber \\
A_{IJK} \!\!\!&=&\!\!\! F^i \partial_i \ln 
(\lambda_{IJK}/Y_{I\bar{I}} Y_{J\bar{J}} Y_{K\bar{K}}), 
\nonumber
\end{eqnarray}
where the indices $i,j$ label the fields with 
nonvanishing F-terms such as the uplifting field $X$ 
and the moduli $T$. 
Here we set $Y_{I\bar{J}}=0$ for $J \ne I$ to reduce the expressions.

\subsection{IIB orientifold model}
A concrete uplifting scheme was first proposed~\cite{Kachru:2003aw} 
in the framework of warped flux compactifications~\cite{Giddings:2001yu}. 
In the type IIB orientifold model, the shape moduli as well as 
the dilaton can be fixed with heavy masses by introducing 
three-form flux in extra dimensions (Calabi-Yau three-folds). 
On top of that, if we consider gaugino condensations on 
D-branes wrapping four-cycles, the corresponding size moduli 
$T$ would receive a superpotential 
$W_{\rm np}(T)=\sum_n^{n_{\rm np}} A_ne^{-a_nf_a(T)}$ 
where $A_n \sim {\cal O}(1)$ and $a_n \sim {\cal O}(4 \pi^2)$. 
Combining this with possible constant contributions induced 
by the three-form flux or nonperturbative effects on three-branes, 
the effective moduli superpotential is given by $W_0=c+W_{\rm np}(T)$. 
Assuming $K_0(T)$, $\partial_T K_0(T)$, 
$\partial_T \partial_{\bar{T}} K_0(T),\ldots={\cal O}(1)$ 
for $T={\cal O}(1)$, the effective $N=1$ moduli potential has only 
SUSY preserving minima with (semi-)negative vacuum energies 
without any fine tuning as mentioned above. 

In order to lift one of the minima to have a vanishing 
vacuum energy and break SUSY, we put an anti D3-brane at 
the tip of a warped throat induced by the three-form flux. 
The anti-brane is sequestered from the size moduli $T$ 
as well as from the MSSM sector due to the warped geometry, 
if we assume, e.g., that the MSSM sector is put far from the 
warped throat in the extra dimensions. In this case, the effect of 
the anti-brane appears just as an uplifting energy which breaks $N=1$ 
SUSY explicitly in the 4D effective theory, and soft terms are 
determined by light moduli F-terms estimated as, e.g., 
$F^T \approx m_{3/2}/\ln (M_{Pl}/m_{3/2})$ for $n_{\rm np}=1$~\cite{Choi:2004sx} 
(see Ref.~\cite{Abe:2005rx} for $l_a \ne 0$). 
If we assume that only the heavy moduli (e.g., shape moduli in this case) 
distinguishes the SM flavors, there would be no dangerous SUSY 
flavor violations~\cite{Choi:2008hn}. 
However, note before mentioning the SUSY flavor violations that, 
it is still challenging to obtain even the SM flavor structure such 
as proper generation numbers and the realistic Yukawa coupling 
matrices within the framework of IIB orientifold models. 

Instead of the anti-brane, we can think of the 
F-term uplifting. We consider the situation that a dynamical 
SUSY breaking occurs somewhere in extra dimensions, which might 
be represented by $X$ at a low energy with the nonvanishing 
Polonyi term $f \ne 0$ in the effective superpotential. 
We can tune the parameters $c$ and $f$ so that the F-term of $X$ 
cancels the negative energy of the moduli potential $W_0$. 
However in this case, direct couplings between $X$ and $Q^I$ 
through $Y_{I\bar{J}}(T,\bar{T},|X|^2)$ could appear in general, 
depending on the explicit construction of $X$ and $Q^I$ sectors. 
Then, the issues of sequestering is model dependent and the 
derivation of $Y_{I\bar{J}}$ model by model might be quite 
complicated with six extra dimensions.

\subsection{5D orbifold model}
The five-dimensional (5D) supergravity provides the simplest 
illustrating framework to study the issues of dynamical uplifting 
and sequestering, because of the singleness of the extra dimension. 
Moreover, it is known that a certain class of realistic Yukawa 
matrices can be realized due to the wavefunction localization 
caused by gauging MSSM matter fields under the graviphoton. 

Thanks to the off-shell dimensional reduction~\cite{Abe:2006eg} which 
is based on the $N=1$ superspace~\cite{Paccetti:2004ri}, we can derive 
$K$, $W$ and $f_a$ in a systematic way for an arbitrary setup of the 
5D supergravity on orbifold $S^1/Z_2$. 
When both $X$ and $Q^I$ originate from 5D hypermultiplets 
charged under the graviphoton, we find $\Omega_0=-3(T+\bar{T})/2$, 
$Y_{X \bar{X}}=(1-|e^{-c_XT}|^2)/c_X$, $L_{IJ}=0$ and 
\begin{eqnarray}
Y_{I \bar{J}} 
\!\!\!&=&\!\!\! 
\Big( \frac{1-|e^{-c_IT}|^2}{c_I}
+\frac{1-|e^{-(c_I+c_X)T}|}{3(c_I+c_X)} |X|^2 
\Big) \delta_{IJ}, 
\nonumber
\end{eqnarray}
where $c_X$ and $c_I$ are the graviphoton charges of $X$ and 
$Q^I$ respectively. The parameters in the superpotential such as 
the Polonyi term $f$ and holomorphic Yukawa couplings $\lambda_{IJK}$ 
appear as constant parameters originating from the superpotential 
at the fixed points of orbifold. The moduli potential $W_0(T)$ 
would be given in the exactly same form as the above IIB model 
by assuming bulk (zero mode) and boundary gaugino condensations. 
Then the argument of the moduli stabilization and uplifting 
is equivalent to the above IIB case. 

An advantage is that we now have exact forms of $Y_{X \bar{X}}$ and 
$Y_{I \bar{J}}$ which are necessary to calculate $m^2_I$ and $A_{IJK}$. 
As mentioned above, we obtain realistic values of physical Yukawa couplings 
$y_{IJK}=\lambda_{IJK}/\sqrt{Y_{I\bar{I}}Y_{J\bar{J}}Y_{K\bar{K}}}$ 
with a certain choice of $c_I$ due to the wavefunction localizations 
(see, e.g., Ref.~\cite{Abe:2004tq} and references therein). 
By using the above soft term formula, we easily find that the contribution 
from $X$ to $m^2_I$ is always tachyonic, and $m^2_I$ and $A_{IJK}$ vanish 
at the tree-level in a large $|c_I|$ (or $|c_X|$) limit with 
$c_Ic_X<0$~\cite{Abe:2008ka}. 
The limit corresponds to the situation that $Q^I$ ($X$) localizes 
more severely toward the opposite fixed point to $X$ ($Q^I$), and 
the sequestering is achieved. 
(In the limit $|c_I|, |c_X| \to \infty$, the fields $Q^I$, $X$, 
respectively, become strictly localized at the fixed point.) 
However, we can also show that the sequestering is incompatible 
with the wavefunction profile that yields realistic Yukawa matrices. 
For the realistic Yukawa couplings, the heavy and the light 
generations need to localize against each other, and then 
either generation is forced to localize toward the SUSY 
breaking field $X$, yielding a large tachyonic $m^2_I$ 
due to the direct coupling. This might be simply because 
we have only single extra dimension.

\section{Summary}
The nonperturbative moduli stabilization with uplifting 
provides a systematic way for realizing a SUSY breaking 
minimum with a vanishing vacuum energy. We have discussed 
the feature of sequestering within such a framework based 
on two illustrating situations. One is a ten-dimensional model 
that would correspond to a certain low energy limit of IIB 
superstring (a top-down approach), and another is the 5D 
orbifold model where all the functions in the low energy 
effective theory are calculable (a bottom-up approach). 

In the former, the realization of SM flavor structure is 
still challenging issue, while in the latter, the 
compatibility between the realistic Yukawa structure 
and sequestering is a nontrivial problem. For IIB models, 
further studies on flavors based on, e.g., magnetized 
extra dimensions~\cite{Cremades:2004wa} would be interesting. 
For 5D models, an extension to the case with multi moduli 
would solve the problem~\cite{Abe:2008an}. Finally, it is 
important to study these issues also in IIA/heterotic 
string models or to extend the 5D model to the case with 
more extra dimensions step by step.

\subsection*{Acknowledgements}
The author would like to thank 
Tetsutaro~Higaki, 
Tatsuo~Kobayashi, 
Yuji~Omura 
and 
Yutaka~Sakamura
for fruitful collaborations. 
The author is supported by the Japan Society for the Promotion 
of Science for Young Scientists (No.182496).

\end{document}